# Single-crystalline high-quality β-Ga$_2$O$_3$ pseudo-substrate on sapphire through sputtering for epitaxial deposition

Guangying Wang[1*], Shuwen Xie[1], William Brand[2], Saleh Ahmed Khan[3], Ahmed Ibreljic[3], Darryl Shima[4], Yueying Ma[1], Brahmani Challa[1], Fikadu Alema[2], Andrei Osinsky[2], Anhar Bhuiyan[3], Ganesh Balakrishnan[4], Shubhra S Pasayat[1].

1. Department of Electrical and Computer Engineering, University of Wisconsin-Madison, Madison, WI 53706, United States of America
2. Agnitron Technology Incorporated, Chanhassen, Minnesota, USA
3. Department of Electrical and Computer Engineering, University of Massachusetts Lowell, Lowell, Massachusetts 01854, USA
4. Center for High Technology Materials, University of New Mexico, Albuquerque, New Mexico 87106

\* . Corresponding author

**Abstract:**
Solid-phase epitaxy (SPE) of β-Ga$_2$O$_3$ thin films by radio-frequency (RF) sputtering and then crystallized through high-temperature post-deposition annealing is employed on sapphire substrates, yielding a high-quality pseudo-substrate for subsequent buffer growth via MOCVD and LPCVD. Low roughness (<0.5 nm) and sharp single-crystalline diffraction peaks corresponding to the (−201), (−402), and (−603) reflections of β-Ga$_2$O$_3$ were observed in the SPE β-Ga$_2$O$_3$ film and the subsequent epitaxial buffer layer. N-doped Ga$_2$O$_3$ film on SPE Ga$_2$O$_3$ film grown by LPCVD showed step-assisted growth mode with reasonable electronic behavior with 45 cm²/V·s mobility at a bulk carrier concentration of $1.3 \times 10^{17}$ cm$^{−3}$. These results suggest that SPE Ga$_2$O$_3$ is a promising pathway to advance the development of β-Ga$_2$O$_3$ on foreign substrates.

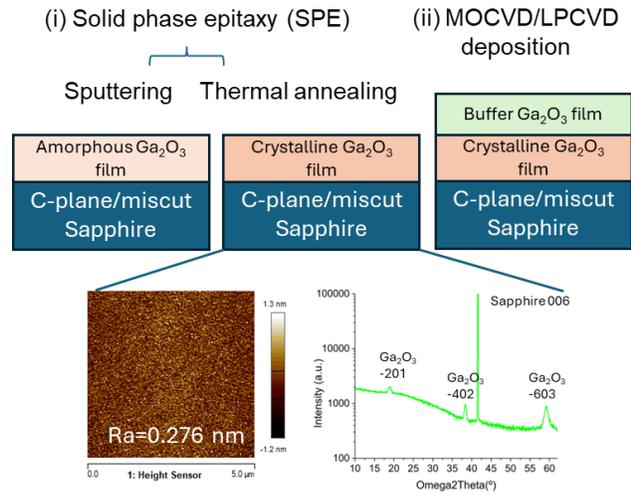

Introduction:
Gallium oxide (Ga$_2$O$_3$) has recently emerged as a highly promising semiconductor for next-generation power electronics. Its ultra-wide bandgap (~4.8 eV), high dielectric constant, and extremely high breakdown electric field (~8 MV/cm) provide the theoretical basis for devices with superior power-handling capabilities compared to conventional wide-bandgap semiconductors such as Silicon Carbide (SiC) and Gallium Nitride (GaN)[1]. These attributes make Ga$_2$O$_3$ attractive for high-voltage transistors, Schottky diodes, and other high-power applications.

Despite these advantages, two fundamental challenges limit their practical deployment. First, the intrinsic thermal conductivity of $Ga_2O_3$ is relatively low (11-27 W·m$^{-1}$·K$^{-1}$, depending on the phase and orientation), which restricts heat dissipation and directly impacts device reliability, efficiency, and lifetime under high-current operation[2]. Second, bulk $Ga_2O_3$ substrates, although commercially available, are expensive, with 4 and 6 inch diameter wafers still under development, posing obstacles for both research scalability and industrial adoption[3–5]. These issues have motivated intensive exploration of thin-film growth on foreign substrates, combining the favorable properties of $Ga_2O_3$ with the thermal and mechanical advantages of scalable platforms.

Several epitaxial techniques have been investigated for this purpose, including Metal–Organic Chemical Vapor Deposition (MOCVD), Molecular Beam Epitaxy (MBE), and Radio-Frequency (RF) Magnetron Sputtering. MOCVD and MBE are capable of producing high-quality homoepitaxial films with well-controlled doping and layer structures. However, when adapted for heteroepitaxy, these methods generally require a nucleation stage to accommodate lattice mismatch, which complicates growth and often results in rough, defect-prone films, particularly at small thicknesses[6]. In contrast, sputtering—a physical vapor deposition technique—offers a simpler and more industrially scalable route for fabricating $Ga_2O_3$ layers. While it is less suited for building complex device heterostructures, sputtering can generate uniform and smooth films with large-area compatibility, making it highly attractive as a foundation for subsequent device fabrication [3,7–10]. This technology has been widely utilized in other III-V group template development such as sputtered AlN as nucleation layer with GaN or AlN buffer layers to create template on sapphire substrates[11–14].

In this work, we propose a two-step pathway to address the material and integration challenges, shown in Figure 1. First, $Ga_2O_3$ thin films are deposited on sapphire substrates by radio-frequency (RF) sputtering and subsequently crystallized through high-temperature post-deposition annealing, yielding epitaxial-quality films via solid-phase epitaxy (SPE). These sputtered and subsequently post-annealed templates serve as high-quality seed layers with improved crystalline order and reduced defect density compared with $Ga_2O_3$ direct growth on sapphire with CVD. In the second step, thicker device-grade $Ga_2O_3$ layers are grown on these SPE templates using MOCVD and Low-Pressure Chemical Vapor Deposition (LPCVD), taking advantage of the improved surface morphology and crystalline quality of the underlying film.

**Experiment:**

For the preparation of SPE β-$Ga_2O_3$ on sapphire substrates, an amorphous 25 nm $Ga_2O_3$ layer was first deposited on both c-plane sapphire and 6° miscut sapphire (toward the m-plane) using on-axis RF magnetron sputtering. The deposition was carried out at a substrate temperature of 200 °C under a chamber pressure of 15 mTorr, with 100 W RF power applied. Crystallization of the sputtered $Ga_2O_3$ was then achieved by annealing in air at 850 °C in a muffle furnace. The resulting crystalline films were subsequently used as templates for buffer growth in MOCVD or LPCVD. For MOCVD-grown undoped β-$Ga_2O_3$ buffers on c-plane sapphire at 820°C, triethylgallium (TEG) was employed as the Ga precursor at a molar flow rate of 9.2 µmol/min under an $N_2$ carrier environment. The VI/III ratio was varied between 1000, 1500, and 3000, while maintaining a film thickness of approximately 200 nm. For the n-type LPCVD β-$Ga_2O_3$ growth, 6° off miscut sapphire is utilized to suppress the island-nucleation-dominated growth mode expected on flat sapphire under high growth rate (6 - 8 µm/hour), where the absence of atomic steps limits ordered adatom incorporation[15]. The deposition was carried out at 1000 °C following the growth conditions described in Reference[16].

Atomic force microscopy (AFM) was carried out using a Bruker Dimension Icon system to examine the surface morphology of the β-Ga$_2$O$_3$ layers. Film thickness of the SPE β-Ga$_2$O$_3$ layers was determined by X-ray reflectivity (XRR), while crystal quality and phase identification were evaluated by high-resolution X-ray diffraction (XRD) ω–2θ scans using a Panalytical Empyrean diffractometer, and scanning transmission electron microscope (HRTEM/STEM) using aberration-corrected Transmission Electron Microscope JEOL NEOARM 200CF. The surface morphology of the LPCVD-grown n-type β-Ga$_2$O$_3$ films on SPE sapphire substrates was further characterized by scanning electron microscopy (SEM). Electrical properties, including electron concentration and mobility, were measured on the same LPCVD-grown films using Hall effect measurements.

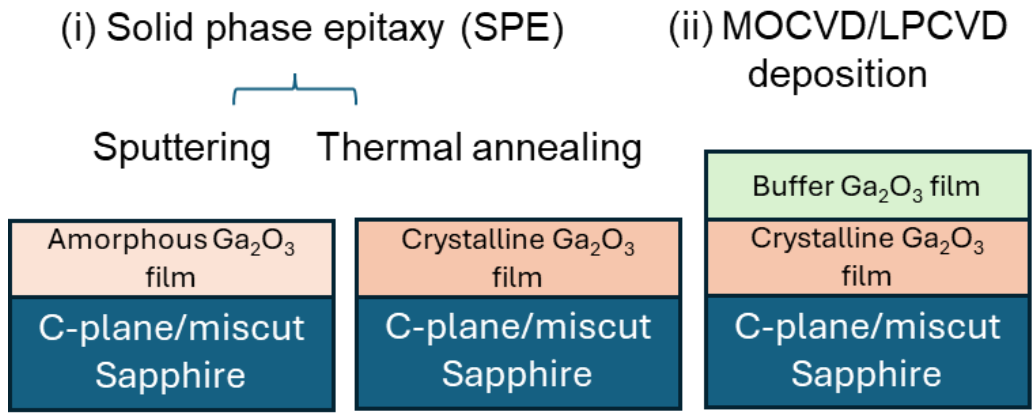

Figure 1. Schematic illustration of the two-step growth pathway: (i) formation of crystalline β-Ga$_2$O$_3$ on sapphire substrates by solid-phase epitaxy (SPE), followed by (ii) deposition of the β-Ga$_2$O$_3$ buffer layer by MOCVD or LPCVD.

Results:

The AFM images of 25-nm SPE crystalline β-Ga$_2$O$_3$ grown on c-plane sapphire substrate and miscut sapphire substrate are shown in Figure 2. For a 1 × 1 μm² scan, the surface roughness was 0.332 nm on the c-plane sapphire substrate and 0.406 nm on the miscut sapphire substrate. At a larger 5 × 5 μm² scan size, the corresponding values were 0.276 nm and 0.356 nm, respectively. Under identical sputtering and post-annealing conditions, the crystalline β-Ga$_2$O$_3$ film on the planar sapphire substrate consistently exhibited lower roughness than the film on the miscut sapphire substrate. This difference is attributed to the higher density of atomic steps on the miscut substrate, which propagate into the epitaxial film and enhance step-related height variations. In both cases, the RMS roughness remained below 0.5 nm, confirming atomically smooth surfaces comparable to those of bulk β-Ga$_2$O$_3$. Moreover, XRD patterns of crystalline β-Ga$_2$O$_3$ films grown on both c-plane sapphire substrate and miscut sapphire substrate exhibited only the (−201), (−402), and (−603) reflections of β-Ga$_2$O$_3$ together with the (006) reflection of sapphire. Annealing was performed at 850 °C, which stabilized the β-Ga$_2$O$_3$ phase. To further verify crystallinity and epitaxial alignment, an off-axis φ-scan of the β-Ga$_2$O$_3$ (−401) reflection was performed, shown in Figure 3. The φ-scan revealed sixfold symmetric peaks, arising from rotational domains induced by the hexagonal symmetry of the sapphire substrate, thereby confirming the epitaxial nature and single-phase β-Ga$_2$O$_3$ growth[17].

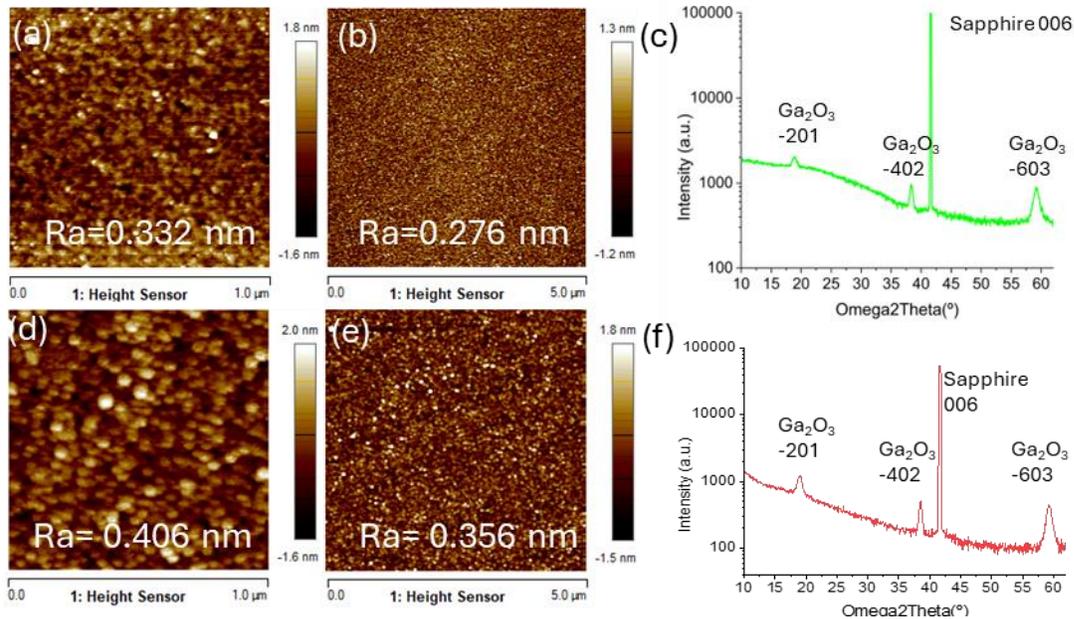

Figure 2. AFM scan of crystalline Ga$_2$O$_3$ on (a) planar sapphire substrate with 1 × 1 μm$^2$ scan size, (b) with 5 × 5 μm$^2$ scan size, (d) 6-degree miscut sapphire substrate with 1 × 1 μm$^2$ scan size, and (e) with 5 × 5 μm$^2$ scan size. XRD diffraction pattern of SPE Ga$_2$O$_3$ on (c) c-sapphire, and (f) miscut sapphire substrate.

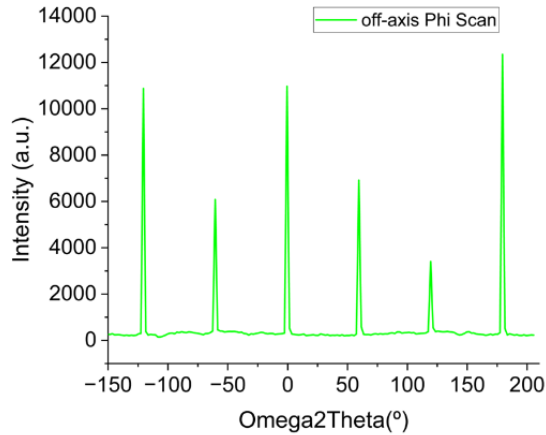

Figure 3. XRD off-axis φ-scan of SPE Ga$_2$O$_3$ on sapphire at β-Ga$_2$O$_3$ (−401).

The SPE-grown β-Ga$_2$O$_3$ layers were subsequently used as substrates for buffer growth. MOCVD was carried out on c-plane sapphire with the SPE β-Ga$_2$O$_3$ layer, while LPCVD was performed on miscut sapphire to take advantage of the faster growth rate achievable on this orientation. For both deposition methods, a bare sapphire substrate without the SPE layer was included as a reference for fair comparison.

Figure 4(a–c) presents the AFM results of ~250-nm-thick β-Ga$_2$O$_3$ buffer layers grown by MOCVD on SPE β-Ga$_2$O$_3$ on c-plane sapphire substrates under different VI/III ratios of 1000, 1500, and 3000. At a scan area of 5 × 5 μm$^2$, the β-Ga$_2$O$_3$ grown directly on bare sapphire substrates exhibited surface roughness values of 3.87 nm, 6.92 nm, and 7.84 nm for VI/III ratios of 1000, 1500, and 3000, respectively. In contrast, when grown on SPE β-Ga$_2$O$_3$ on c-plane sapphire substrates, the roughness was consistently reduced to 3.74 nm, 4.73 nm, and 5.50 nm for the same

VI/III ratios. These results indicate that buffer layers grown on SPE templates suppress facet development, particularly under lower VI/III ratios. VI/III ratios of 1000 showed the lowest surface roughness among all three VI/III conditions, possibly due to reduced gas phase reaction[18,19]. Across all three conditions, MOCVD-grown β-$Ga_2O_3$ buffers on SPE sapphire substrates consistently demonstrated lower surface roughness compared to those on bare sapphire substrates. The structural quality was further confirmed by XRD analysis, shown in Figure 5. Films grown on SPE β-$Ga_2O_3$/c-plane sapphire exhibited sharp diffraction peaks corresponding only to the (−201), (−402), (−603), and (−804) planes of β-$Ga_2O_3$, together with the (006) reflection of sapphire. In contrast, MOCVD-grown β-$Ga_2O_3$ on bare sapphire showed additional parasitic peaks, indicative of secondary phases or misoriented grains.

Furthermore, TEM study was performed on $Ga_2O_3$ on SPE $Ga_2O_3$ on sapphire sample, shown in Figure 6. Clear lattice orientation can be observed both at the SPE $Ga_2O_3$ to sapphire interface and MOCVD $Ga_2O_3$ to SPE $Ga_2O_3$ interface, indicating the preferred -201 orientation of the $Ga_2O_3$ film. However, defects are also observed in Figure 6 (b) in the SPE $Ga_2O_3$ film, indicating further optimization is needed to improve uniformity. These findings in XRD and TEM demonstrate that the SPE β-$Ga_2O_3$ template provides a clear advantage in both crystalline quality and surface morphology compared to direct growth on bare sapphire substrates.

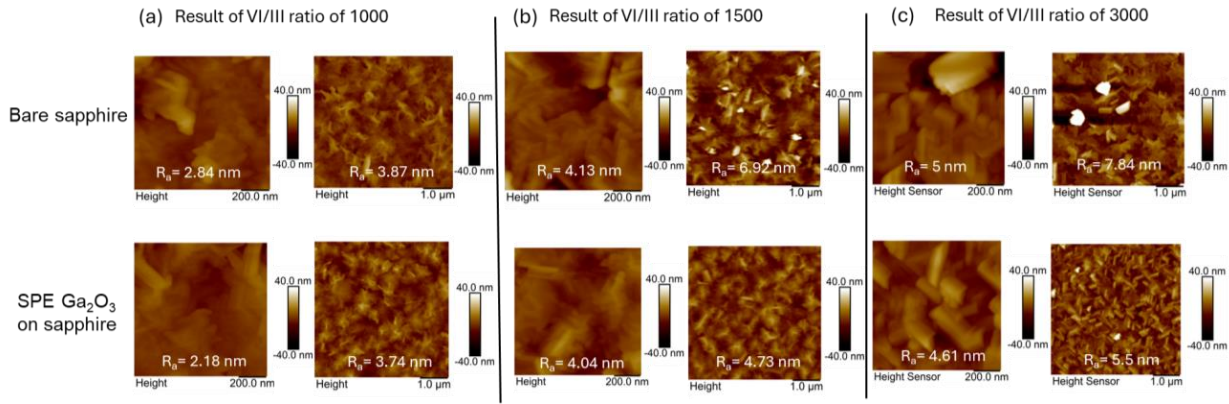

Figure 4. AFM result of MOCVD grown $Ga_2O_3$ with VI/III ratio of (a) 1000, (b) 1500 and (c) 3000.

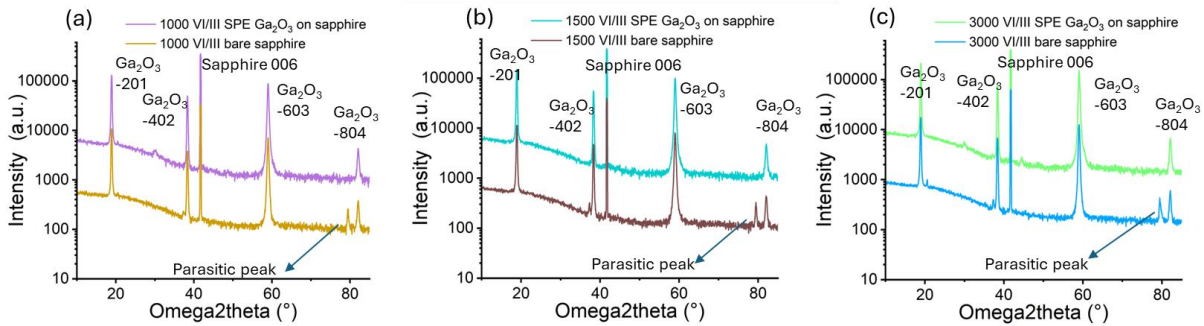

Figure 5. XRD result of MOCVD grown $Ga_2O_3$ with VI/III ratio (with an offset in intensity to show the parasitic peak) of (a) 1000, (b) 1500 and (c) 3000.

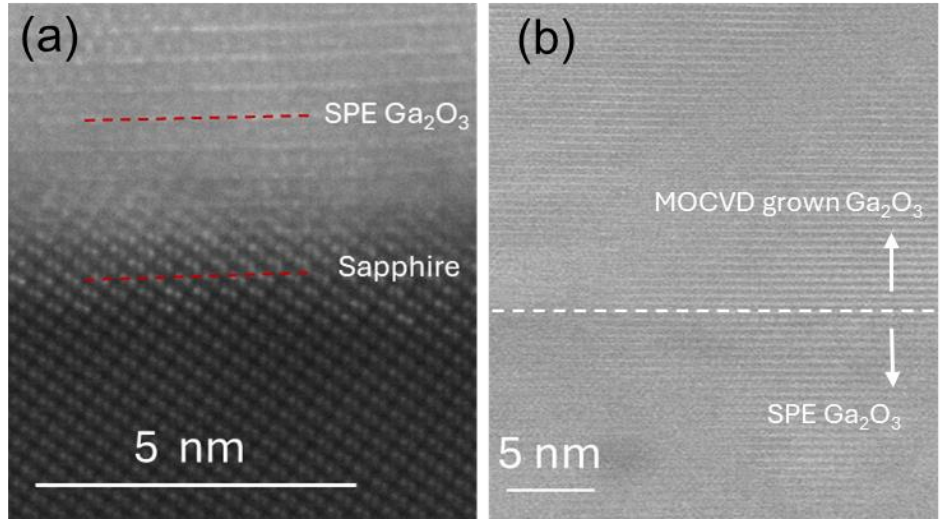

Figure 6. TEM result of MOCVD grown $Ga_2O_3$ on SPE $Ga_2O_3$ on sapphire with VI/III ratio of 1000 at the (a) SPE $Ga_2O_3$ and sapphire interface with red dashed line showing lattice orientation matching, (b) MOCVD $Ga_2O_3$ on SPE $Ga_2O_3$ on sapphire with white dashed line showing the $Ga_2O_3$ interface.

The Field Emission Scanning Electron Microscopy (FESEM) based characterization of the surface morphology and electrical transport characteristics result of the LPCVD β-$Ga_2O_3$ films using Hall measurements, deposited on SPE $Ga_2O_3$ on 6° off-axis sapphire is shown in Figure 7 (a). The film exhibited elongated surface features aligned with the off-cut direction, which is characteristic of a step-assisted deposition mode. This behavior has been reported previously for LPCVD β-$Ga_2O_3$ on vicinal sapphire [15,16], where the increased density of atomic steps suppress random island nucleation, reduce rotational domain formation, and support more coherent lateral ordering. SPE pseudo-substrate sample followed the expected deposition behavior on the vicinal surface and yielded a mobility of 45 cm²/V·s at a bulk carrier concentration of $1.3 \times 10^{17}$ cm⁻³, which aligned with the transport behavior associated with step-flow-facilitated deposition on 6° off-axis sapphire. The XRD results, shown in Figure 7 (b), further support these observations. The LPCVD-grown β-$Ga_2O_3$ film on bare miscut sapphire displayed parasitic peaks, whereas the film on the SPE template exhibited dominant β-$Ga_2O_3$ (−201), (−402), (−603), and (−804) reflections along with the sapphire (006) peak. For the film on the SPE template, additional weak polycrystalline peaks were detected, consistent with the SEM morphology that revealed faceting and mixed-phase regions. However, these peaks were of very low intensity compared to the dominant (−201) reflection, indicating that they originate from a thin fraction near the film surface. Further LPCVD optimization is planned to suppress these features and achieve uniform single-phase films, in order to further improve the film morphology and carrier mobility.

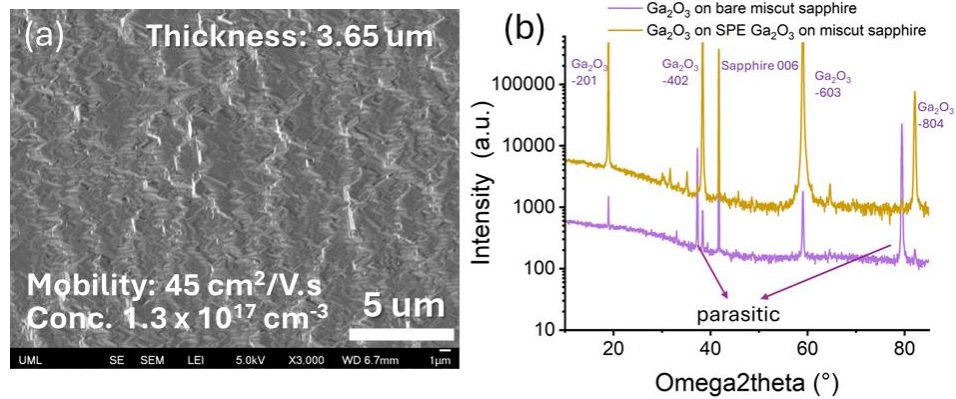

Figure 7. Surface morphology and transport properties of LPCVD β-Ga$_2$O$_3$ grown films on SPE-β-Ga$_2$O$_3$ template on 6° off miscut sapphire, and (b) XRD result of LPCVD grown Ga$_2$O$_3$ on SPE Ga$_2$O$_3$ on miscut sapphire substrate and on bare miscut sapphire substrate.

Conclusion:

In this work, we demonstrated the deposition of single-crystalline SPE β-Ga$_2$O$_3$ films on both c-plane and miscut sapphire substrates. The SPE β-Ga$_2$O$_3$ films exhibited sub-nanometer roughness (<0.5 nm) and sharp single-crystalline diffraction peaks corresponding to the (−201), (−402), and (−603) reflections of β-Ga$_2$O$_3$. Furthermore, both MOCVD-grown and LPCVD-grown β-Ga$_2$O$_3$ films deposited on SPE β-Ga$_2$O$_3$ on sapphire showed a clear reduction in parasitic peaks compared with reference films grown on co-loaded sapphire substrates. LPCVD-grown n-type β-Ga$_2$O$_3$ films on SPE β-Ga$_2$O$_3$ on sapphire exhibited step-assisted deposition mode with reasonable electron mobility and electron concentration.

By integrating physical vapor deposition and chemical vapor deposition, our approach combines the advantages of the high scalability and smooth surface by sputtering with the precise compositional control and superior electrical quality offered by CVD. As a result, we realize single-crystalline β-Ga$_2$O$_3$ films with excellent surface smoothness, strong structural integrity, and enhanced electrical performance. This combined strategy provides a promising pathway to advance the development of β-Ga$_2$O$_3$ for both fundamental research and practical device applications.

**Corresponding Author**
Guangying Wang - Electrical & Computer Engineering, University of Wisconsin-Madison, Madison, Wisconsin 53706, United States; Email: gwang265@wisc.edu
**Authors**
Shuwen Xie, Yueying Ma, Brahmani Challa, Shubhra S. Pasayat
Department of Electrical and Computer Engineering, University of Wisconsin–Madison, Madison, Wisconsin 53706, United States
William Brand, Fikadu Alema, Andrei Osinsky
Agnitron Technology Inc., Chanhassen, Minnesota 55317, United States
Saleh Ahmed Khan, Ahmed Ibreljic, Anhar Bhuiyan
Department of Electrical and Computer Engineering, University of Massachusetts Lowell, Lowell, Massachusetts 01854, United States

Darryl Shima, Ganesh Balakrishnan
Center for High Technology Materials, University of New Mexico, Albuquerque, New Mexico 87106, United States

**Notes**
The authors declare no competing financial interest.

Acknowledgment:

This project was funded by the Air Force Research Lab Midwest Hub.

The authors gratefully acknowledge the use of facilities and instrumentation in the UW-Madison Wisconsin Center for Nanoscale Technology. The Center (wcnt.wisc.edu) is partially supported by the Wisconsin Materials Research Science and Engineering Center (NSF DMR-2309000) and the University of Wisconsin-Madison.

S. A. Khan, A. Ibreljic, and A. Bhuiyan acknowledge funding support from the National Science Foundation under Award Nos. ECCS-2501623 and ECCS-2532898.

Electron microscopy was carried out in the Nanomaterials Characterization Facility at the University of New Mexico, a facility that is supported by the State of New Mexico, the National Science Foundation and the National Aeronautics and Space Administration. The acquisition of the JEOL NEOARM AC-S/TEM at the University of New Mexico was supported by NSF grant DMR-1828731 and NASA Emerging Worlds grant 80NSSC21K1757